\documentclass{jetpl}
\twocolumn

\usepackage[russian]{babel}
\usepackage{graphicx}
\usepackage[pdfborder={0 0 0}, colorlinks=true, linkcolor=blue, citecolor=blue, urlcolor=blue]{hyperref}

\begin{document}

%%% article in English
\lat

%%% article title
\title{Quantitative comparison of LDA+DMFT and ARPES spectral functions}

%%% article title - for colontitle (at the top of the page)
\rtitle{Quantitative comparison of LDA+DMFT and ARPES spectral functions}

%%% article title - for table of contents (usualy identical with \title)
\sodtitle{Quantitative comparison of LDA+DMFT and ARPES spectral functions}

%%% author(s) ( + e-mail)
\author{I.\,A.\,Nekrasov\/\thanks{e-mail: nekrasov@iep.uran.ru},
N.\,S.\,Pavlov\/\thanks{e-mail: pavlov@iep.uran.ru}}

%%% author(s) - for colontitle (at the top of the page)
\rauthor{I.\,A.\,Nekrasov, N.\,S.\,Pavlov}

%%% author(s) - for table of contents
\sodauthor{Nekrasov, Pavlov}

%%% author's address(es)
\address{Institute of Electrophysics, Ural Branch of the Russian Academy of Sciences, 620016 Ekaterinburg, Russia}

%%% dates of submition & resubmition (if submitted once, second argument is *)
%\dates{}{*}

%%% abstract
\abstract{
The emergence of angle-resolved photoemission spectroscopy (ARPES) made it possible to observe electronic dispersion directly as a spectral function map. On the other hand, a spectral function map can be obtained theoretically, for example, in the LDA+DMFT method. The electronic band on such a map is characterized not only by its energy position at a given $k$-point, but also by its width and intensity. To illustrate a way of quantitative comparison of theoretical spectral functions and ARPES data, spectral functions obtained by the LDA+DMFT method are chosen. It is shown that the theoretical spectral functions should take into account a number of experimental features: the photoionization cross section, the experimental energy and angular resolution, as well as the effects of the photohole lifetime arising in the process of photoemission. In this article, we present a robust procedure for taking these experimental features into account by the example of iron-based high-temperature superconductors (HTSC) systems: NaFeAs and FeSe on a SrTiO$_3$ substrate.
}

%%% PACS numbers

\maketitle

\textbf{1. Introduction.}
In the last few decades, the intensive development of angle-resolved photoemission spectroscopy (ARPES), in particular, associated with the discovery of copper high-temperature superconductors (HTSC) due to their quasi-two-dimensionality, made it possible to observe experimentally the electron band structure for various classes of materials with a fairly good instrumental resolution and in a quite wide range of binding energies~\cite{Damascelli_ARPES_2003}, including  weakly correlated three-dimensional systems (see $e.g.$ Ref.~\cite{ARPES_SnAs}). The corresponding ARPES data are maps on which the electronic states at a given point in reciprocal space are characterized by energy position, width and intensity.

On the other hand, the improvement of theoretical methods for calculating electronic states, for example, density functional theory (DFT) and its combination with methods that take into account electronic (for example, LDA+DMFT~\cite{LDADMFT_Kotliar2006}) or any other interactions, also makes it possible to obtain the electronic band structure in the form of spectral function maps. This has caused an urgent need for a quantitative comparison of theoretical and experimental electronic spectral function maps. For this, in the theoretical and experimental data it is necessary to compare not only the qualitative energy position of the electronic quasiparticle bands, but also their relative intensities and widths.

In this work, it is shown that an important factor, influencing the width and intensity of the experimentally observed electron bands, to a large extent is the conditions under which an experiment was carried out: the incident beam  energy and the instrumental resolution. In addition, there are certainly temperature broadening and damping associated with electron-electron (or some other) interactions in the system under study. The incident beam energy determines the relative probability of a given electronic state to be excited~\cite{Damascelli_ARPES_2003,Damascelli_ARPES_2013}. This significantly affects the relative intensity of various branches of the electronic spectrum on a spectral function map. Also, the experimental instrumental resolution (angle and energy resolutions) provides a sizable contribution to the width of the electron bands ~\cite{Damascelli_ARPES_2013}. Another remarkable contribution to the width and intensity of the electronic spectrum (which is not directly related to the experimental setup) is the finite lifetime of a photo-hole arising in the process of photoemission~\cite{Damascelli_ARPES_2003}.

Here we propose how to introduce into theoretical spectral function map a number of experimental features: the photoionization cross section, experimental energy and angular resolutions, as well as the effects of the photo-hole lifetime. By the example of iron based HTSC (NaFeAs and FeSe on a SrTiO$_3$ substrate), it is shown that a significant contribution to the broadening of quasiparticle bands is associated precisely with taking these experimental details into account within the theory. It is also shown that the choice of the parameters of the broadening of the initial theoretical bands can lead to a change in the energy position of the maxima of the spectral function. In addition, the issue of the need to introduce some alphanumeric standard for the designation of the intensity scale is discussed. This is important for quantitative comparative analysis of both experimental and theoretical spectral function maps from various sources.

\begin{figure*}[h]
	\center{\includegraphics[width=1.0\linewidth]{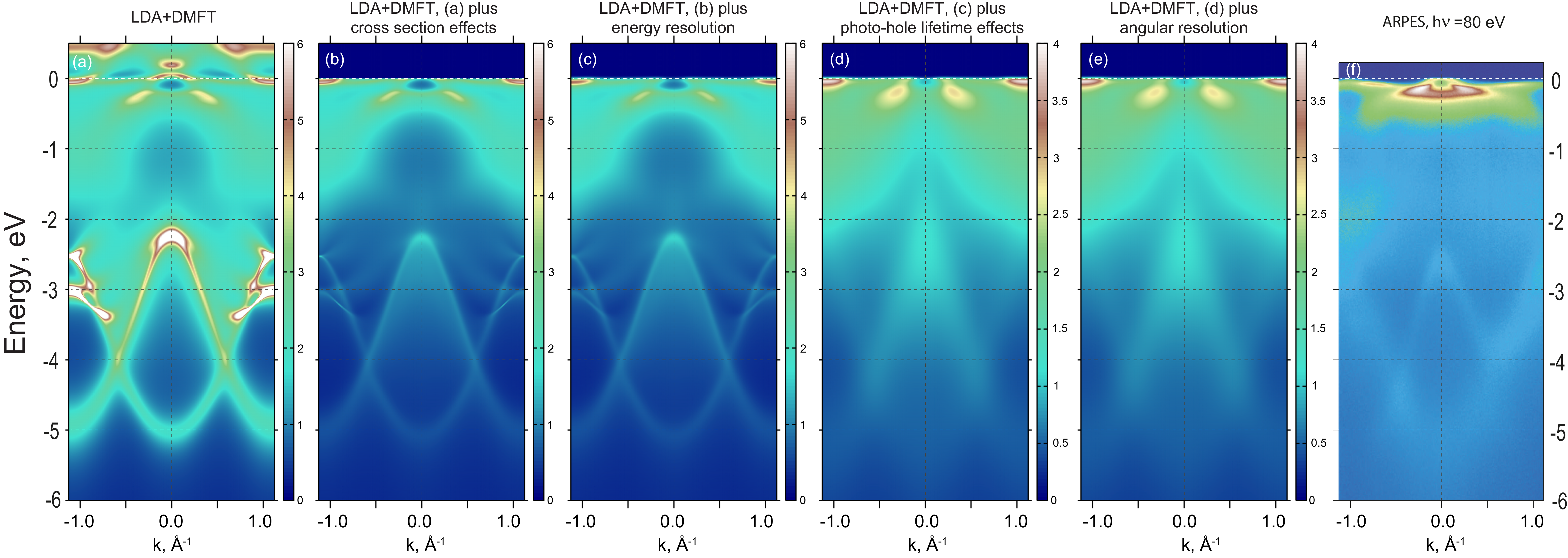}}
	\caption{Fig. 1. (Color online) Panel (a) - LDA+DMFT spectral function for NaFeAs in the M-$\Gamma$-M direction~\cite{NaFeAs}. Panel (b) - taking into account the photoionization cross section. In panel (c), in addition to (b), the experimental energy resolution is taken into account. Then on panel (d) additionally the effect of the photo-hole lifetime is considered. Further on panel (e), the angular experimental resolution is taken into account. Panel (f) -- ARPES data from~\cite{NaFeAs_ARPES}. Fermi level -- zero energy (white dotted line).}
	\label{NaFeAs_broaden_80}
\end{figure*}

\textbf{2. Results.}
To study the effect of taking into account experimental features on the spectral function map, we used the data obtained earlier in LDA+DMFT(CT-QMC) calculations for superconductors NaFeAs~\cite{NaFeAs} and FeSe on a SrTiO$_3$ substrate~\cite{Nekrasov_FeSe3} (for details of calculations and description of methods, see the works above). The Fig.~\ref{NaFeAs_broaden_80}(a) shows the LDA+DMFT spectral function map in the high symmetry M-$\Gamma$-M direction for the NaFeAs superconductor.

On Fig.~\ref{NaFeAs_broaden_80}(b) the LDA+DMFT spectral function with the photoionization cross section included is plotted. To obtain the valence band, the LDA+DMFT spectra were convoluted with the Fermi distribution function at the experimental temperature $20$~K. In this ARPES experiment (Fig.~\ref{NaFeAs_broaden_80}(f))~\cite{NaFeAs_ARPES}, the incident beam energy was $80$~eV, which corresponds to the relative values of the photoionization cross section in the atomic limit Fe-3d:As-4p $\ =6.181:0.08177$~\cite{Yeh_Lindau_1,Yeh_Lindau_2}. At the given incident beam energies, due to the photoionization cross section, the intensity of the Fe-3d states significantly increases in comparison with As-4p.

Then, to take into account the experimental energy resolution of $20$~meV, the convolution of the spectral function (Fig.~\ref{NaFeAs_broaden_80}(b)) with the Gaussian function was performed. To this end the so-called {\it energy distribution curves} (EDC) are used here for convolution (Fig.~\ref{NaFeAs_broaden_80}(c)). The effect of account of the energy resolution is hardly noticeable in the LDA+DMFT spectral function map, since the electron-electron correlations themselves give rise to broadening and suppression of the intensity of quasiparticle bands, comparable in magnitude with the experimental resolution.

In our previous work~\cite{Sr2RuO4_broadening_Gaussian_2007} (see also papers cited there) it was shown, that in case of binding energy dependence of the photohole lifetime, arising due to process of photoemission, theoretical $k$-integrated photoemission spectra better agrees with experimental ones. In this work, we have applied this approach to the spectral function by performing the convolution with a Gaussian function with a width at half maximum which is linearly dependent on the binding energy $C\cdot\epsilon_B+\Gamma_{exp}$. Here, $\epsilon_B$ is the binding energy, $\Gamma_{exp}$ is the experimental energy resolution, and $C$ characterizes the decrease in the photo-hole lifetime when moving away from the Fermi level deep into the valence band (see Refs.~\cite{Sr2RuO4_broadening_Gaussian_2007,broadening_Lorentzian_1992,broadening_Gaussian_2000}). The maximum value of the half-width of the Gaussian function was chosen 0.5~eV at energy -1~eV and below it remains constant. Thus obtained spectral function map is shown in Fig.~\ref{NaFeAs_broaden_80}(d). Account of the finite lifetime of a photo-hole leads to a serious damping and suppression of quasiparticle bands intensity, making them practically indistinguishable. The maximum intensity value on the spectral function map has been reduced from 6 to 4 in order to highlight its structure (see intensity color scale).

The final step for a quantitative comparison with ARPES was accounting for the experimental angular resolution -- $0.5$ degrees (in the $k$-space $0.04$ \AA$^{-1}$, correspondingly). It was implemented via the convolution of the spectral function with the Gaussian function. The so-called {\it momentum distribution curves} (MDC) are used here for convolution (Fig.~\ref{NaFeAs_broaden_80}(e)).

The above analysis makes it possible to distinguish the broadening and damping of quasiparticle bands, caused by electron correlations (or other interactions) in the system, from the broadening and damping of the spectral function associated with the experimental setup and the process of photoemission itself.

We see that introduction of the experimental features for the LDA+DMFT spectral function leads to good agreement with the experimental data (Figs.~\ref{NaFeAs_broaden_80}(e) and~\ref{NaFeAs_broaden_80}(f)). In particular, for NaFeAs electron-electron correlations are moderately manifested (mass renormalization at the Fermi level is about 3) and damping of the LDA+DMFT spectral function is insignificant (Fig.~\ref{NaFeAs_broaden_80}(a)). While in the experiment (Fig.~\ref{NaFeAs_broaden_80}(f)), a rather diffuse intense formation is observed near the Fermi level. Taking into account the above-mentioned experimental features, the damping of the LDA+DMFT spectral function significantly increases, which is in good agreement with experiment. It is also worth noting that the LDA+DMFT spectral function below $-2$~eV, formed mainly with As-4p states, together with the experimental features also better agrees with the ARPES data in terms of structure and intensity.

The next system considered in this work is the FeSe superconductor on the SrTiO$_3$ substrate (FeSe/STO). A distinctive feature of this system is the presence in the ARPES spectra of the so-called ``shallow'' and twin bands around 50 meV below the Fermi level near the M-point~\cite{FeSe_ARPES,Nekrasov_FeSe}. Thus, using the FeSe/STO system as an example, one can see the influence of accounting for experimental features on a small energy scale.
\begin{figure*}[h]
	\center{\includegraphics[width=1.0\linewidth]{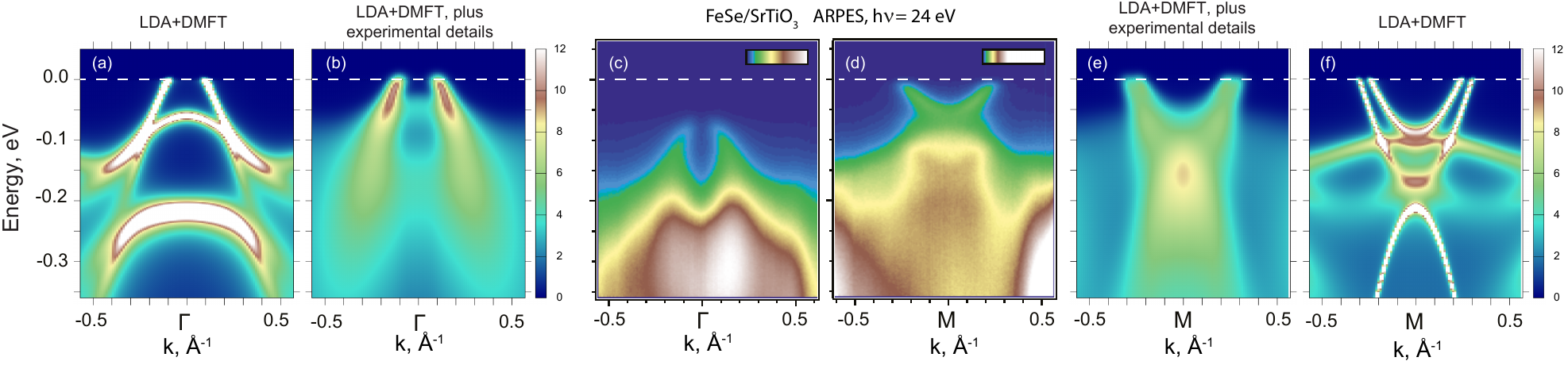}}
	\caption{Fig. 2. (Color online) Panels (a), (f) -- LDA+DMFT spectral function for FeSe/STO in the M-$\Gamma$-M and $\Gamma$-M-$\Gamma$ directions~\cite{Nekrasov_FeSe3}. On panels (b), (e) the experimental features indicated in the text are taken into account. Panels (c), (d) -- ARPES from~\cite{FeSe_ARPES}. Fermi level -- zero energy (white dotted line).}
	\label{FeSe_broaden}
\end{figure*}
\begin{figure*}[h]
	\center{\includegraphics[width=0.7\linewidth]{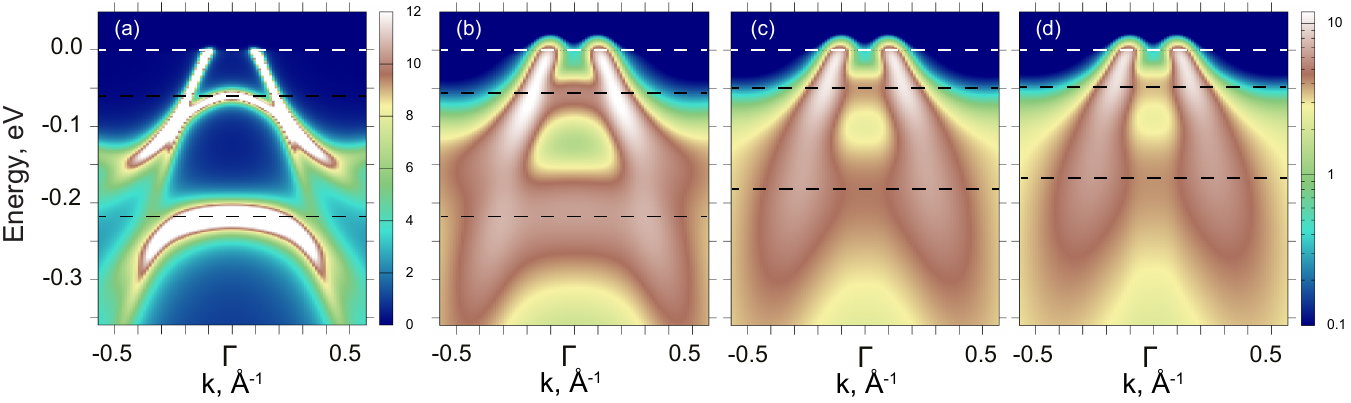}}
	\caption{Fig. 3. (Color online) Dependence of the position of the maximum of the intense region on the rate of change in the photo-hole lifetime with increasing binding energy. Panel (a) -- LDA+DMFT spectral function for FeSe/STO in the  M-$\Gamma$-M direction (linear intensity scale). On panels (b)-(d) the experimental features indicated in the text for different lifetimes of the photo-hole are taken into account: (b) $0.192\cdot\epsilon_B+0.008$; (c) $0.392\cdot\epsilon_B+0.008$; (d) - $0.492\cdot\epsilon_B+0.008$ (logarithmic intensity scale). The horizontal dashed lines show the position of the intensity maximum of the corresponding bands at the $\Gamma$-point. Fermi level -- zero energy.}
	\label{FeSe_broaden_lin}
\end{figure*}

The Fig.~\ref{FeSe_broaden} shows a serial introduction of experimental features described above for FeSe/STO when going from LDA+DMFT quasiparticle bands to ARPES spectra. In the case of FeSe/STO, the energy of incident photons in the ARPES experiment is $24$~eV~\cite{FeSe_ARPES}, which gives the photoionization cross section in the atomic limit e-3d:Se-4p $\ =5.656:5.626 $  -- practically $1:1$. Note that for the energy $24$~eV these values of the photoionization cross section were obtained by linear interpolation between the available energies in the tabular data of Refs.~\cite{Yeh_Lindau_1,Yeh_Lindau_2}. Other experimental characteristics: the energy resolution is 8 meV, the angular resolution is 0.3 degrees (in  $k$-space  $0.025$ \ \AA$^{-1}$, correspondingly), the temperature of the experiment is $16$~K.

In this paper, we will not discuss in detail the problem of the presence in the LDA+DMFT of a hole pocket at the $\Gamma$--point, which is not observed experimentally. We only note that this can be caused by the suppression of the intensity of the Fe-3d$_{xy}$ orbital in the given geometry of the experiment, as well as by the influence of the STO substrate and the method of sample growth (doping, distortion, etc.) (for more details, see~\cite{Nekrasov_FeSe}). Similar LDA+DMFT results with a hole pocket at the $\Gamma$-point were obtained in ~\cite{Haule_FeSeSTO}, however, without direct comparison with ARPES data.

It can be noted that taking into account the experimental features gives a quantitatively similar structure of the LDA+DMFT and ARPES data except for the bands at the Fermi level near the $\Gamma$-point. Thus, LDA+DMFT quasiparticle bands become less pronounced and much more diffuse. In this case, the ``shallow'' band ($\sim~50$~meV) near the M-point, obtained in the LDA+DMFT calculation, acquires a quantitatively similar structure (like a “tank-top”) in comparison with ARPES (see Fig.~\ref{FeSe_broaden} panels (d) and (e)).

Especially we would like to discuss shift of the intensity maxima of the spectral function depending on how fast the photo-hole lifetime grows with binding energy increase. This shift can be of the order of 100 meV. It is interesting that in this case the initial LDA+DMFT bands do not change its energy position. The position of the quasiparticle bands is determined by the experimentators by the intensity maximum of the spectral function. Thus energy position of experimental quasiparticle band may differ from its true position, since the speed of growth of the photo-hole lifetime with binding energy is not known in practice. This can especially strongly affect the interpretation of experimental data on small energy scales (up to 1 eV) below the Fermi level.

The shift of the most intense region near the $\Gamma$-point for different values of speed of growth of the photo-hole lifetime from 0 (Fig.~\ref{FeSe_broaden_lin}(a) to $\sim~0.5$ (Fig.~\ref{FeSe_broaden_lin}(d)) is shown.  For the band near $-0.22$~eV the shift is of the order of $50$~meV, and for the band near $-0.06$~eV -- $15$~meV. This shift of the intense region might be important for the analysis of ARPES data on such small scales, especially for superconducting systems. Since the effects of the photo-hole lifetime are always present, we can conclude that the energy position of the quasiparticle bands in ARPES is determined with an accuracy up to the speed of growth of the photo-hole lifetime with binding energy increase. For the experimental data analyzed in this work, the magnitude of the shift can be greater than the instrumental resolution  ($8$ and $20$ meV, respectively). This fact should be borne in mind when analyzing and interpreting ARPES spectra.

An important parameter that affects the visual perception of the spectral function and the analysis of its structure is the intensity of the spectral function map and the choice of the method to display it. The intensity scale is often shown by changing a certain set of colors, where a certain intensity value corresponds to a certain color from the palette. Thus, the display of the intensity scale of the spectral function map can vary in the maximum and minimum cutoff values and in the ``distance'' between adjacent colors, which allows one to emphasize the necessary ARPES data details. However, for a quantitative comparison of ARPES with theoretical spectra, it would be good to have the same or similar in color and magnitude scale of the intensity of the spectral function map. To do this, it is necessary to know the color scale of ARPES spectra, which is often given, and also the value of the spectral function corresponding to a given color, which is practically never presented by both experimentators and theoretitians.

Another ``arbitrary parameter'' is the choice of linear or logarithmic intensity scale. The latter is usually used to highlight the low-intensity parts of the spectral function map. Figs.~\ref{FeSe_broaden}(b) and~\ref{FeSe_broaden_lin}(d) show the same spectral function in one color scale, but in different intensity scales: linear and logarithmic.

For convenience of quantitative comparison of data from different sources, one can designate the corresponding scale in some alphanumeric way, for example, lin0N3T4.5G7Y8Br10W (see Fig.~\ref{FeSe_broaden_lin}(а)), which means a linear scale, 0 - Navy, 3 - Turquoise, 4.5 - Green, 7 - Yellow, 8 - Brown, 10 - White, or simply should be described it in the text of the article.

\textbf{3. Conclusion.}
In this work, using the example of LDA+DMFT spectral functions, it is shown that for a quantitative comparison of theoretical spectral function maps with ARPES data, it is necessary to take into account experimental features: the photoemission cross section, experimental energy and angular resolution, as well as the effects of the photo-hole lifetime associated with the process of photoemission itself. Note that account of these experimental features is not related to the model concept of photoemission and is necessary in the analysis of theoretical results. It is also shown that the interpretation of ARPES spectra is complicated by the presence of a photo-hole lifetime depending on the binding energy (which explicit form is not known for a particular material), which leads to a shift in the intensity maxima of spectral functions relative to the true position of the quasiparticle band. The importance of a detailed description of the color intensity scale for the quantitative comparison of spectral maps from different sources is shown.

\textbf{Acknowledgments.}
This work was carried out with partial support from the Russian Foundation for Basic Research (grant 20-02-00011). The work of N.S. Pavlov was partially supported by the grant of the President of the Russian Federation MK-1683.2019.2.
The calculations were performed on the ``URAN'' supercomputer of the Institute of Mechanics and Mathematics, Ural Branch of the Russian Academy of Sciences.

\bibliographystyle{jetpl}

\end{document}